# Surface Instabilities and Patterning in Liquids: Exemplifications of the "Hairy Ball Theorem"

Edward Bormashenko

*Ariel University, Physics Faculty, 40700, P.O.B. 3, Ariel, Israel*

**Abstract**

Application of the "hairy ball theorem" to the analysis of the surface instabilities inherent for liquid/vapor interfaces is reported. When a continuous tangential velocity field exists on the surface of the liquid sample which is homeomorphic to a ball, zero velocity points will be necessarily present at the surface. The theorem is exemplified with the analysis of the instability occurring under the rapid evaporation of polymer solutions. Zero velocity points, accumulating pores, enable direct visualization of the instability. The patterning may be essentially different on the surface of a torus.

## 1. Introduction

Surface instabilities, arising as a result of a variety of physico-chemical factors, play a tremendous role in nature and technology [1-3]. The classical types of these instabilities for liquid/vapour interfaces were studied more than one hundred years ago by Rayleigh and Plateau [4-5]. Lord Rayleigh and Plateau showed that infinite liquid cylindrical shapes are inherently energetically unstable when perturbed at sufficiently large wavelengths, and will evolve to form arrays of drops [4-6]. Thin liquid films also may destabilized under the influence of an external field, which can simply be gravitational. This destabilization occurs under Rayleigh–Taylor instability [6-7]. The Rayleigh-Plateau and Rayleigh–Taylor instabilities result in a certain equilibrium state, characterized by the static liquid/vapour interface.

In parallel, there exist dynamic instabilities, under which the particles constituting the liquid/air interface are moving continuously. The typical representative of these instabilities is the instability arising from the Benard-Marangoni convection, shown in **Fig. 1.** In the presence of temperature (or concentration) gradients on a liquid–fluid interface, the temperature (or the concentration of the solute) dependence of surface tension causes the tangential movement of a liquid [8-12]. Surface-tension-driven Benard-Marangoni convection in liquid layers heated from below exhibits an instability, leading to a diversity of physical phenomena, including patterning in the evaporated polymer solutions [13-18]. This kind of patterning gives rise to large scale patterns, such as depicted in **Fig. 2.** The mechanism responsible for the large-scale patterning observed in rapidly evaporated polymer solutions remains debatable [19]. In spite of the usual attribution of patterning to the temperature-gradient-driven Marangoni instability, it was shown recently that heating from below destroys the pattern [19]. Thus, the patterning may be related to solute-capillary Marangoni flow, or to the instability introduced by de Gennes [20]. It should be emphasized that in all kinds of aforementioned instabilities the tangential vector field of velocities, drives the liquid. The present paper focuses on the topological aspects of patterning, caused by surface instabilities.

**Results and discussions.**

Consider patterning, caused by surface instabilities from the point of view of the "hairy ball theorem". The hairy ball theorem of algebraic topology states that there is no non-vanishing continuous tangent vector field on even-dimensional $n$-spheres [21]. The simpler (and less general) wording of this theorem states that any continuous tangent vector film on the sphere must have at least one point where the vector is zero. The witty exemplification of this remarkable topological theorem may be formulated in a following way: "if a sphere is covered in hair and we try to smoothly brush those hair to make them all lie flat, we will always leave behind at least one hair standing up straight or a hole" [22].

Till now applications of the "hairy ball theorem" to physical problems remain scanty [23-25].We demonstrate, that the "hairy ball theorem" is applicable to the analysis of the dynamic surface instabilities. Consider the surface instability resulting from the Benard-Marangoni instability, shown in **Fig. 1**. The pattern arising from this instability may be seen as a set of *N* elementary cells depicted in **Fig. 1**. In every cell the tangential non-zero vector field is defined on a surface of a cell. The cell (or a number of cells) is homeomorphic to a ball (the number of cells *N* does not matter).

Thus, according to the "hairy ball theorem" there exists at least one point at which the velocity is zero (these points for the Benard-Marangoni cells are shown in **Fig. 1**).

In the experiments performed with rapidly evaporated polymer solutions zero-velocity points allow effective visualization of the instability, appearing when the solution is cooled from above, as a result of evaporation [13-15]. Solid particles or pores (zero mass particles), accumulated in zero-velocity points, make the surface pattern, due to instability, visible, as shown in **Fig. 2**. Pores, accumulated in zero-velocity points separate cells formed under rapid evaporation of polymer solutions. Optical microscopy enables observation of the movement of pores towards zero-velocity boundaries as discussed in Ref.14 and shown in **Fig. 3**.

**Fig. 1** depicts the situation when in one of the sub-cells, resulting from the Benard-Marangoni instability, a liquid rotates clockwise, whereas in the second one it rotates counterclockwise. Remarkably, the "hairy ball theorem" predicts the existence of at least one zero velocity point at the surface of the liquid also in this particular case and explains, why we do not observe a sole cell under the Benard-Marangoni convection, but a number of cells. Obviously, the characteristic dimensions of cells result from the thermodynamic and kinetic considerations.

**Conclusions**

Our topological analysis of the patterns arising from surface instabilities was exemplified with the study of the pattern, arising from the instability occurring under rapid evaporation of polymer solutions. However, the same analysis may be undertaken for other dynamic instabilities, occurring at the liquid/vapor interface, when the tangential velocity field is defined on the whole surface of the sample. In all cases when the sample is topologically equivalent (homeomorphic) to a ball, at least one zero-velocity point necessarily exists on the surface, according to the "hairy ball theorem". These points may allow direct visualization of the instability, accumulating solid tracers or pores, as it takes place for evaporated polymer solutions.

**Acknowledgements**

I am thankful to Mrs. Y. Bormashenko for her kind help in preparing this manuscript. I am indebted to Professor A. Blumberg (Austin University, USA) and Professor Gene Whyman (Ariel University, Israel) for inspiring discussions.

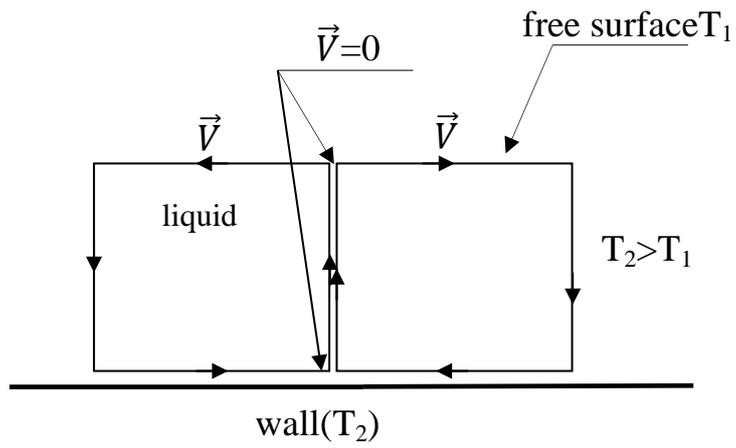

**Fig. 1.** Formation of Benard-Marangoni cells. Liquid is heated from below. Particles constituting the free surface of a liquid are moving tangentially. Zero-velocity points are shown.

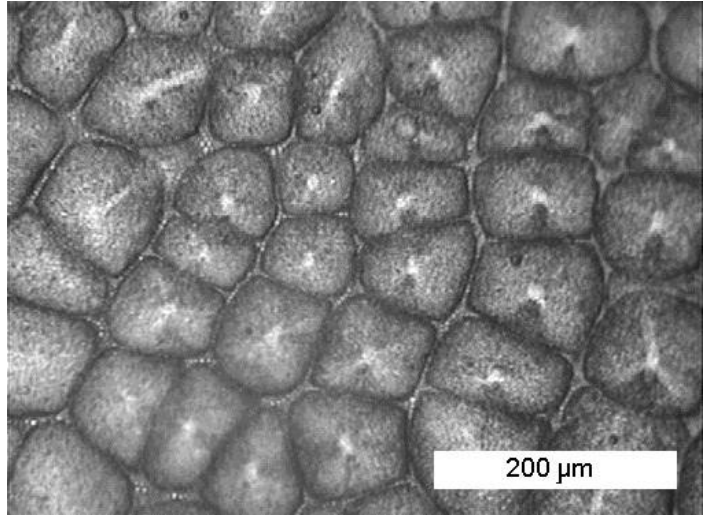

**Fig. 2a**. The pattern observed under evaporation of polycarbonate dissolved in dichloromethane (7% wt.).

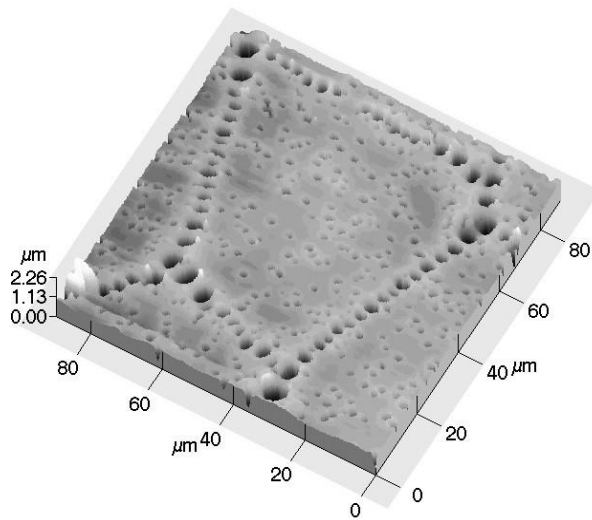

**Fig. 2b**. Typical boundary separating cells formed under rapid evaporation of polymer solutions. AFM image of the boundary, observed for polymethyl(methacrylate) dissolved in chloroform (10% wt.). The boundary is built from microscopically scaled pores.

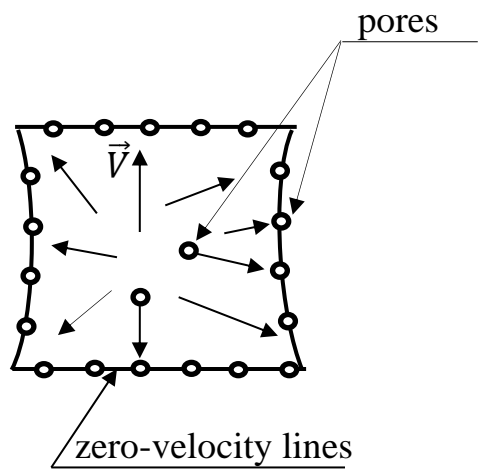

**Fig. 3**. Velocity field observed at the interface of the evaporated polymer solution (Ref. 14). Pores are accumulated at the zero-velocity lines.